\documentclass[prl,twocolumn,aps,superscriptaddress,showpacs,preprintnumbers]{revtex4}

\usepackage{varioref,exscale,latexsym,amsmath,amssymb}
\usepackage{graphicx}
\usepackage{bm}
\usepackage{slashed}
\usepackage[colorlinks=true, pdfstartview=FitV, linkcolor=black, citecolor=black, urlcolor=black]{hyperref}

\begin{document}

\newcommand{\be}{\begin{equation}}
\newcommand{\ee}{\end{equation}}
\newcommand{\beq}{\begin{eqnarray}}
\newcommand{\eeq}{\end{eqnarray}}
\newcommand{\bea}{\begin{eqnarray}}
\newcommand{\eea}{\end{eqnarray}}
\newcommand{\beqn}{\begin{eqnarray}}
\newcommand{\eeqn}{\end{eqnarray}}
\newcommand{\rd}{\mathrm{d}}
\newcommand{\X}{\mathbb{X}}
\newcommand{\ack}[1]{{\bf Pfft! #1}}
\newcommand{\pa}{\partial}
\newcommand{\osigma}{\overline{\sigma}}
\newcommand{\orho}{\overline{\rho}}
\newcommand{\myfig}[3]{
	\begin{figure}[ht]
	\centering
	\includegraphics[width=#2cm]{#1}\caption{#3}\label{fig:#1}
	\end{figure}
	}
\newcommand{\littlefig}[2]{
	\includegraphics[width=#2cm]{#1}}
\newcommand{\1}{{\rm 1\hspace*{-0.4ex}%
\rule{0.1ex}{1.52ex}\hspace*{0.2ex}}}

\def\pa{\partial}
\def\dd{\!\cdot \!}
\def\S{\mathbb{S}}
\def\pp{\cal{P}}

\title{Born Reciprocity in String Theory and the Nature of Spacetime}

\author{Laurent Freidel}\email{lfreidel@perimeterinstitute.ca}
\affiliation{ Perimeter Institute for Theoretical Physics, 31 Caroline St, N, On N2L 2Y5, Waterloo, Canada}

\author{Robert G. Leigh}\email{rgleigh@uiuc.edu}
\affiliation{Department of Physics, University of Illinois, 1110 West Green St., Urbana, IL, 61801, U.S.A}

\author{Djordje Minic}\email{dminic@vt.edu}
\affiliation{Department of Physics, Virginia Tech, Blacksburg, VA 24061, U.S.A.}

\date{\today}

\begin{abstract}
After many years, the deep nature of spacetime in string theory remains an enigma. In this letter we incorporate the concept of Born reciprocity in order to provide a new point of view on
string theory in which spacetime is a derived dynamical concept. This viewpoint may be thought of as a dynamical chiral  phase space formulation of string theory, in which Born reciprocity is implemented as a choice of a Lagrangian submanifold of the phase space, and amounts to a generalization of T-duality. In this approach the fundamental symmetry of string theory contains phase space diffeomorphism invariance and the underlying string geometry should be understood in terms of dynamical bi-Lagrangian manifolds and an apparently new geometric structure, somewhat reminiscent of para-quaternionic geometry, which we call Born geometry.
\end{abstract}

\pacs{11.25.-w,11.25.Pm,11.25.Hf}

\maketitle
String theory is a remarkable model that aims to be a description of the quantum nature of spacetime. 
Yet the true nature of spacetime in string theory is still rather mysterious. In this letter we present a new interpretation of string theory based on the concept of Born reciprocity \cite{born} which elucidates this fundamental question. 

The Born reciprocity principle states that the validity of quantum mechanics implies 
a fundamental symmetry between space and momentum space. This symmetry results from the freedom to choose a basis  of  states.
General relativity fundamentally breaks this symmetry because it states that spacetime is curved, while energy-momentum space, defined as a cotangent space, is linear and flat.
The simple but  radical idea proposed by Max Born more than 75 years ago \cite{born}, is that in order to unify quantum mechanics and general relativity one should also allow phase space, and thus momentum space, to carry curvature \cite{dms}.
Up to now, however, the mathematical implementation and the physics of Born geometry have been elusive \cite{bornstring, veneziano}.
In this letter we show that Born geometry naturally appears in the very foundations of string theory and that it underlies many exotic stringy space-time properties.
In the standard formulation of perturbative string theory, there exist many signs of novel structures that should appear at short spacetime distances. Perhaps one of the simplest is the concept of T-duality on flat compact target spaces,  one of the hallmarks of perturbative string theory \cite{tdual}. This concept is central in the study of fixed angle, high energy scattering in string theory \cite{grossm} (including its generalizations \cite{prlwitten}), the study of the high temperature limit \cite{witten}, and the still mysterious stringy uncertainty principle \cite{grossm,sup}. 
In the open string sector, T-duality played a fundamental role in the discovery of D-branes \cite{dbrane}. Mirror symmetry can be also viewed as T-duality \cite{syz}. What these and other studies make clear is that the short distance behavior of string theory is exotic, at least from the perspective of quantum field theory. In the case of T-duality on flat compact target spaces, the short distance behavior is governed by long distance behavior in some dual space. In this letter we ask: are there similar conclusions that can be reached in more general settings, for instance, when target space is curved or non-compact? Is the usual foundational assumption, that string perturbation theory is built on maps from the worldsheet into a smooth spacetime, truly justified?
In what follows, we re-consider some of these basic assumptions, and reformulate string theory in a larger context. We will emphasize that some of the structure of traditional string perturbation theory is dictated not by general principles of quantization and consistency, but by auxiliary ad hoc requirements, including locality. 
Relaxing these auxiliary requirements and letting the string take its fullest extension will allow a reformulation that implements quantum mechanical Born reciprocity.
In so doing, T-duality is cast essentially as a Fourier transform. In fact, many of the concepts that we discuss in this letter are familiar, at least in the context of a $\sigma$-model with fixed flat compact target. One of our motivations is to understand how to generalize the usual picture of T-duality on compact target spaces \cite{tdualnew} to non-compact and curved cases, as well to generalize recent 
efforts on the geometric underpinnings of the traditional picture found in the context of double field theory \cite{double}. However, we also introduce new concepts associated with the diffeomorphism symmetry in phase space
as well as the mathematical structures of bi-Lagrangians and Born geometry which we believe incorporate the main features of stringy spacetime.
In particular, the new concept of Born geometry discussed in this letter contains the traditional picture of T-duality as well as the results of double field theory as important special cases of a more general structure.

{\it Quasi-Periodicity and generalized T-duality:}
The simplest examples of T-duality arise by considering string theory in flat backgrounds. Thus, we begin the discussion by examining the quantization of the Polyakov action coupled to a flat metric,
\beq
\frac{1}{2\pi \alpha' } 
S_{P}(X) =\frac{1}{4\pi \alpha'}\int_{\Sigma} \eta_{\mu\nu} (*\rd X^{\mu} \wedge  \rd X^{\nu}),
\eeq
where $*,d$ denote the Hodge dual and exterior derivative on the worldsheet, respectively.
We generally will refer to local coordinates on $\Sigma$ as $\sigma,\tau$, while it is traditional to interpret $X^\mu$ as local coordinates on a target space $M$, here with Minkowski metric. 
One needs to demand that the integrand be single-valued on $\Sigma$. For example, on the cylinder $(\sigma,\tau)\in [0,2\pi]\times [0,1]$ this implies
that $\rd X^{\mu}(\sigma,\tau)$ is periodic  with respect to $\sigma$ with period $2 \pi$. 
However, this does not mean that $X^{\mu}(\sigma,\tau)$ has to be a periodic function, even if $M$ is non-compact.
Instead,  it  means that $X^{\mu}$ must be a {\it quasi-periodic} function which satisfies 
$
X^{\mu}(\sigma + 2\pi, \tau) = X^{\mu}(\sigma,\tau) + \bar{p}^{\mu}.
$
Here $\bar{p}^{\mu}$ is the quasi-period of $X^{\mu}$.
If $\bar{p}^{\mu}$ is not zero, there is no {\it a priori} geometrical interpretation of a closed string propagating in a flat spacetime -- periodicity goes hand-in-hand with a spacetime interpretation. Of course, if $M$ were compact and spacelike \cite{nb}, then $\bar p^\mu$ would be interpreted as winding, and it is not in general zero. 

In what follows we will see that the string can be understood more generally to propagate inside a portion of a phase space. 
What matters here is not that string theory possesses or not a geometrical interpretation but whether it can be defined consistently. This is no different than the usual CFT perspective, in which there are only a few conditions coming from quantization that must be imposed; a realization of a target spacetime is another independent concept.  It has always been clear that the concept of T-duality must change our perspective on spacetime, including the cherished concept of locality, and so it is natural to seek a relaxation of the spacetime assumption.

The first hint that it is consistent to consider the more general class of quasi-periodic boundary conditions comes about as follows.
Given a boundary $\pa \Sigma$ parameterized by $\sigma$, a string state $|\Psi\rangle$ may be represented by a Polyakov path integral
$\Psi[x^{a}(\sigma)] =\int_{X|_{\partial \Sigma}=x} [DX] \int_{Met(\Sigma)}[ {\cal D} \gamma]\ e^{\frac{i}{2\pi\alpha'}S_P(X)},$
where $[{\cal D} \gamma]$ denotes the integration measure over the space of 2d metrics.
We begin with the usual assumption that the fields $X$ are periodic, that is $\oint_{C} \rd X =0 $, for any closed loop $C$ on $\Sigma$.
Such a loop carries momentum $\alpha' p= \oint_{C}*\rd X$.
We define a Fourier  transform of this state 
by 
$\tilde{\Psi}[y(\sigma)] \equiv \int [Dx(\sigma)]\, e^{\frac{1}{2\pi i}\int_{\pa\Sigma} x^\mu \rd y_\mu
} \Psi[x^{\mu}(\sigma)].$
In fact,  this state can also be represented as a  string state associated 
to a dual Polyakov action, by extending $y(\sigma)$ to the bulk of the worldsheet, and interpreting $\int_{\pa\Sigma}xdy=\int_\Sigma dX\wedge dY$.
Integrating out $X$ then gives
$\tilde{\Psi}[y(\sigma)] = \int_{Y|_{\partial \Sigma}=y} [DY] \int_{Met(\Sigma)}[ {\cal D} \gamma]\  
e^{-\frac{i\alpha'}{2\pi}S_P(Y)}.$
 The momentum may now be expressed as $p= \oint_{C}\rd Y$, and so we will refer to $Y$ as coordinates in momentum space.
The key difference however compared to the previous path integral 
is that this integral is over quasi-periodic $Y$, as the quasi-period is just $p$. Moreover these quasi-periodic functions are constrained to carry no dual-momenta:
$
\bar{p} =-\alpha'\oint * \rd Y=0.
$
Thus, we see that it is a matter of convention that we have taken $\bar p$ to vanish. Indeed, in the compact case, the Fourier transform is just implementing the T-duality \cite{tdualnew}, and in that case it is well-known that the boundary conditions can be relaxed to finite $(p,\bar p)$. 
{\it However, the notion that T-duality could be viewed as a Fourier transform is much more general and it can be applied in
non-compact and curved cases as well.}

Relaxing the boundary conditions has the following effects. Consider the string path integral with insertions of vertex operators $\sim \prod_{i} e^{ip_{i}X(z_{i})}$. Each of these operators induces multivaluedness in momentum space, with periods $p_{i}$ around each puncture.
This can be rewritten in terms of {\it dual vertex operators} that are non-local operators on the worldsheet, $\sim e^{i \sum_{i} p_{i} \int_{e_{i}} * \rd Y}.$
It turns out that the expectation value of the vertex operator in spacetime is equal to the expectation value of the dual vertex operator in momentum space. Moreover the effect of the dual vertex operator is to open up the momentum space string and allow for monodromies $p_{i}$ around the punctures. 
Note that the 2d electrostatic picture of the correlation functions of the usual vertex operators \cite{grossm} (see also \cite{witten}) now generalizes to 2d electromagnetism with electric and magnetic charges (dyons), if we allow vertices with both $p,\bar p$. {\it The corresponding Dirac-Schwinger-Zwanziger quantization of the dyon charges is equivalent to satisfying the diffeomorphism constraint in the presence of these operators} \cite{flmlong}. Thus, although we have given up (temporarily, as it will turn out) spacetime locality and mutual locality on the worldsheet ({\it i.e.}, absence of branch cuts in the operator product algebra of dyonic vertex operators), the string path integral can still be consistent if the target space of the $\sigma$-model is not itself physical spacetime. In what follows we will construct just such a $\sigma$-model.



{\it First order formalism and the phase space action:}
In view of the preceding discussion consider the first order action
\beq\label{PXactionFlat}
\hat{S} = \int_{\Sigma} \left( \bm{P}_\mu \wedge\rd X^\mu+ \frac{\alpha'}{2} \eta^{\mu\nu} (*\bm{P}_\mu \wedge \bm{P}_\nu) \right).
\eeq
Here $\bm{P}_{\mu}=P_{\mu}\rd\tau + Q_{\mu}\rd \sigma$ is a one-form, and the momentum carried by a closed  loop $C$ on the worldsheet is given by 
$p_{\mu}=\int_{C} \bm{P}_{\mu}$. 
If we integrate out $\bm{P}_\mu$, we find
$*\bm{P}_\mu=\frac{1}{\alpha'} \eta_{\mu\nu} \rd X^\nu
$
and we obtain the Polyakov action plus a boundary term which is exactly the kernel of the above Fourier transform,
$
{\hat{S}} =-\frac{1}{\alpha'} S_{P}(X)+ \int_{\pa\Sigma} X^\mu  \bm{P}_\mu.
$
On the other hand, if we integrate out $X$ instead, we get
$\label{intXflat}
\rd \bm{P}_\mu=0,$ and so we can locally write 
 $
 \bm{P}_\mu= \rd Y_\mu.
 $
 It is in this sense that there is ``one degree of freedom" in  $\bm{P}$ -- on-shell $\bm{P}$ is equivalent to the scalar $Y$.
 Notice though that this is true only locally, and in order to interpret it globally we must allow
 $Y_{\mu}$ to be multi-valued on the worldsheet. That is, $Y_{\mu}$ should carry, as compared to $X^{\mu}$, additional monodromies associated with each non-trivial cycle of $\Sigma$.
 This means that the function $Y$ is only quasi-periodic with periods given by
 $
 \int_{C} \bm{P}_{\mu} = \int_{C}\rd Y_{\mu}=p_{\mu}.
 $
 The action of $Y$ becomes  essentially the Polyakov action
 $
 \hat{S} =-\alpha' S_{P}(Y).
 $
This action is weighted with $\alpha'$ instead of $1/\alpha'$ because $Y$ naturally lives in momentum space.

Starting from (\ref{PXactionFlat})
we see that if we integrate the one form  $\bm{P}$ we get back the spacetime Polyakov action and if we integrate $X$ we get the momentum space Polyakov action. In order to get a phase space action a natural idea is to partially integrate out $\bm{P}$.  
Given the natural worldsheet space and time decomposition of $\bm P_{\mu}= P_{\mu} \rd \tau + Q_{\mu}\rd \sigma $ the action reads (in conformal co-ordinates \cite{footnote1})
$
\hat{S} = \int  P_{\mu} \partial_{\tau}{X}^{\mu} -  Q_{\mu} \partial_{\sigma}X^{\mu} + \frac{\alpha'}{2} (Q_{\mu}Q^{\mu}- P_{\mu}P^{\mu} ).
$
The equations of motion (EOM) for $P,Q$ are simply  
$
\alpha' P = \partial_{\tau}{X}$,
$ \alpha' Q= \partial_{\sigma}X.
$
By integrating $Q$ only,  we get the action in the Hamiltonian form:
$
\hat{S}=\int P\cdot\pa_\tau X -\left(\frac{\alpha'}{2} P\cdot P +\frac1{2\alpha'} \partial_{\sigma}X\cdot \partial_{\sigma}X \right).
$
Now, as suggested by the preceding discussion, we can introduce a momentum space coordinate $Y$ such that 
$\partial_{\sigma} Y= P$. Like $X$, this coordinate is not periodic; its quasi-period $ Y(2\pi) -Y(0) $  represents the string momentum. Using this coordinate the action becomes 
$
\int \partial_{\sigma}Y\cdot \partial_{\tau}{X} - \frac12 \left( \alpha' \partial_{\sigma}Y\cdot \partial_{\sigma}Y + \frac1{\alpha'} \partial_{\sigma}X \cdot \partial_{\sigma}X
\right).$
The main point is that in this action both $X$ and $Y$ are taken to be quasi-periodic.
The usual Polyakov formulation is recovered if one insists that $X$ is single-valued, and the usual T-duality formulation is recovered if one insists that quasi-periods of $X$ appear only along space-like directions and have only discrete values.

We can now embark on several levels of generalization. First, one can assume that  the background metric $G_{\mu\nu}$ and axion $B_{\mu\nu}$ are constant but arbitrary.
In order to express the result it  is convenient, as suggested by the double field formalism \cite{double}, 
to introduce  coordinates $\X^{A}\equiv (X^{\mu}/\sqrt{\alpha'} ,Y_{\mu}\sqrt{\alpha'} )^{T}$  on phase space $\cal P$,
together with two metrics: 
$H_{AB} \equiv   \left( \begin{array}{cc} (G- BG^{-1}B) & (BG^{-1}) \\ -(G^{-1}B) & G^{-1}   \end{array} \right)
$ and  $\eta_{AB} = \left( \begin{array}{cc} 0 & 1 \\ 1 & 0  \end{array} \right)$. $\eta$ is a  neutral metric  and $H$ a generalized Lorentzian one
\cite{signature}. 
These data are not independent: if we define $J\equiv \eta^{-1}H$, then $J$  is an involutive  transformation preserving $\eta$,
that is, $J^{2}=1$, and  $\,\, J^{T}\eta J=\eta.$ 
We call  $(\eta,J)$ a {\it chiral } structure on $\cal P$, with generalized metric $H=\eta J$.
The phase space action is then
\be \label{Tseyt}
S= \int \frac12 (\partial_{\tau}{\X}^{A} \partial_{\sigma}\X^{B} \eta_{AB}  - \partial_{\sigma}\X^{A} \partial_{\sigma}\X^{B} H_{AB}).
\ee
In honour of its inventor \cite{reviews},
we call this the Tseytlin action \cite{chiral}.
If $\eta$ is constant, $J$ depends only on $X$ along non-compact directions, 
quasi-periods are only along $Y$, except when there are compact flat directions, and this action 
is equivalent to the Polyakov action on an arbitrary curved non-compact manifold or 
on a flat torus bundle over a non-compact curved space. This is the realm of double field theory \cite{double}.
However, note that if we allow for `dyonic' vertices, that is vertex operators constructed as functions of $\X$, we expect that the $\sigma$-model can be relaxed away from constant $\eta$ and $H$. Given the interpretation of $X$ and $Y$, this means that not only will space-time become curved, but momentum space as well. This then will lead to an implementation of Born reciprocity.

Indeed, we now propose three layers of generalization.
First we allow $H$ to depend on both $X,Y$ irrespective of whether the directions are flat or non-compact. Second we 
allow non-trivial quasi-periods both along $X$ and $Y$ even when they label non-compact or timelike directions. 
And finally, we relax the condition on $\eta$ to be a flat metric, i.e. we allow it to be arbitrary curved.
These generalizations restore the Born duality symmetry. Of course, it is not clear that these generalizations are consistent, and we will look for consistency conditions coming from the quantum dynamics of the string \cite{flmlong1}.

The dynamics of the sigma model is characterized by this  action
together with a set of four constraints: the Weyl ($W$) and Lorentz ($L$) constraints, expressing the invariance under local world sheet Weyl rescaling and Lorentz transformations, together with the Hamiltonian ($H$) and diffeomorphism ($D$) constraints expressing invariance under 2d diffeomorphisms \cite{flmlong}. In phase space terms, these constraints are
 \be
 \begin{array}{rcl ccr} 
 \quad &\quad W = 0,& &\quad H =   \frac12 \partial_{\sigma}\X\dd J (\partial_{\sigma}\X),&\\
& \,\,\,L = \frac12 \S\dd \S, &
& \!\!D =  \partial_{\sigma}\X\dd \partial_{\sigma}\X,&
 \end{array}
 \ee
 where $\cdot$ denotes the $\eta$ contraction, and we have introduced the vector
 $
 \S^{A} \equiv \partial_{\tau} \X^{A} - J^{A}{}_{B}  \partial_{\sigma} \X^{B}
 $. 
 We now see that in this setting we have to cancel two anomalies: Weyl and  Lorentz.
 These equations express a relaxation of the locality equations $\S=0$ to  milder conditions.
 We also see that in this formalism $H$ and $D$ are on the same footing and should be treated similarly. Finally, in the phase space covariant approach one can  study the correlation functions of dyonic vertex operators that satisfy these constraints \cite{flmlong}.
In the case in which  all metrics are constant, the EOM one gets by varying $\X$ gives 
$\partial_{\sigma} \S^{A}  =0 $.
Up to a time-dependent redefinition $\X^{A} \to \X^{A} + \mathbb{C}^{A}(t)$, this implies the duality equation
$ *\rd X = \alpha' \rd Y$!
In turn this duality equation implies the EOM for $X,Y$, 
$
\Delta \X^{A}=0.
$
This is one of the main points that we stress: {\it the worldsheet EOM is simply a self-duality equation in phase space, and
this fact generalizes to curved backgrounds in phase space.}


We now look at the
 EOM for the sigma model (\ref{Tseyt}) in which $H$ and $\eta$ are no longer constrained to be flat.
 Since we have two metrics on phase space we can consider several different connections: for example, we denote by $\nabla$  the  torsionless connection \cite{connection}
 compatible with the neutral metric $\eta$, while $D$ denotes the one compatible with the generalized metric $H$.
With the help of these and the vector $\S$ we can write the EOM as \cite{flmlong} 
\be\label{eom1}\boxed{
\nabla_{\sigma}\S_{A} = -\frac12 (\nabla_{A}H_{BC})\partial_{\sigma}\X^{B} \partial_{\sigma}\X^{C}.}
\ee
We see that whenever $H$ is not covariantly constant with respect to the $\eta$-compatible connection, we no longer have that 
$\S$ vanishes. Instead, eq. (\ref{eom1}) describes how $\S$ changes along the string.
The equation of motion of the string should be supplemented with the above constraints.
In particular, the Lorentz and diffeomorphism constraints imply that $\S$ and $\pa_\sigma\X$ are null with respect to $\eta$.
The condition that $\S$ is null then implies
$(\nabla_{\S}H_{BC})(\partial_{\sigma}\X)^{B} (\partial_{\sigma}\X)^{C} =0,$
where $\nabla_{\S}$ denotes the derivative along $\S$.
Note also that we can write the eq. (\ref{eom1}) in an alternative form,
$
\nabla_{\sigma}\pa_{\tau}\X =  J (D_{\sigma}\pa_{\sigma}\X )
.$
The RHS denotes the acceleration of the curve $\X(\sigma)$ in the geometry of $H$. It vanishes when it is geodesic. The LHS denotes the rate of change in time of the velocity vector along the curve.
One might interpret this form of the EOM to mean that the geometry ``viewed'' by $\pa_{\sigma}\X$ is  $H$, while the one ``viewed'' by $\pa_{\tau}\X$ is $\eta$.

{\it Born Geometry and bi-Lagrangians:}
The Tseytlin action depends on a choice of a {\it chiral structure} $(\eta,J)$, on $\cal P$,
{\it i.e.}, a neutral metric $\eta$ and an involution  $J$ preserving  $\eta$,
which in turn, allows the construction of a generalized metric $H\equiv \eta J $.
In order to solve the equation of motion we need a 
{\it bi-Lagrangian} structure \cite{bl} compatible with $(\eta,J)$: that  is, a choice of decomposition of $T{\cal P} = L\oplus \tilde{L}$ in terms of two distributions $L,\tilde{L}$ which are null with respect to $\eta$ and such that $J(L)=\tilde{L}$. 
Equivalently, such a bi-Lagrangian  is characterized by an involutive map $K$ which anti-commutes with $J$ and with $\eta$: 
$K^{2}=1$, while  $KJ+JK=0$ and $K^{T}\eta K =- \eta$.
This map is defined by $K|_L=\mathrm{Id}$, $K|_{\tilde{L}}=-\mathrm{Id}$.
We call a manifold $\cal P$ equipped with a chiral and compatible bi-Lagrangian structure $(\eta,J,K)$ a {\it Born manifold}
if $L$ and $\tilde{L}$ are involutive.
Quite remarkably, a Born manifold is equipped with a {\it symplectic} structure on $\cal P$ given by $\omega \equiv \eta K$. It is also equipped with an almost {\it K\"ahler} structure $I\equiv KJ $, $I^{2}=-1$, $I^{T}\omega I = \omega  $  such that the corresponding K\"ahler metric is the generalized metric $ H= \omega I$.
In summary, a Born manifold is a phase space equipped with a symplectic form $\omega$ and metric $H=\omega I$,  that is almost K\"ahler ($I^{2}=-1$, $I^{T}\omega I =\omega$) . It is chiral ($J^{2}=1$, $J^{T}\omega J =\omega$) and it is
bi-Lagrangian (or para-K\"ahler) ($K^{2}=1$, $K^{T}\omega K =-\omega$) and it is equipped with a neutral metric $\eta=\omega K$.
The three structures $(I,J,K)$ anti-commute with each other.
There is no standard nomenclature for this type of geometry \cite{three} and thus we call it {\it Born geometry}.
This new geometric structure naturally unifies the complex, real and symplectic geometries encountered in quantum theory, general relativity and the Hamiltonian formulation of classical theory \cite{gibbons}.

The  Lagrangian distribution $L$ is a generalization of the concept of spacetime and the restriction of the generalized metric to one Lagrangian is 
 the generalization of the concept of spacetime metric: $H|_L \equiv G$. 
 We say that the bi-Lagrangian distribution $L,\tilde{L}$  is {\it transversal } with respect to the chiral structure if
the metric on $L$ is  covariantly constant along $\tilde{L}$, that is
$
\nabla_{\tilde{U}}G =0,$ for $
\tilde{U} \in \tilde{L}.
$
One can show \cite{flmlong} that solutions of the classical string EOM associated with a chiral structure $(\eta,J)$ on $\cal P$
are in  correspondence with transversal bi-Lagrangian distributions.
In the flat case, $\S=0$, and if $\pa_\sigma\X$ is in $L$, then $\pa_\tau \X$ is in $\tilde L$, because  $ \pa_{\tau}\X= J(\pa_{\sigma} \X) $ when $\S=0$ and because $J: L\to \tilde L$. Thus, in the flat case, the spacetime in which string propagates can be identified with $L$. In the general case, we have seen that the Lorentz and diffeomorphism  constraints imply that $\S$ and
$\pa_\sigma \X$ are null with respect to $\eta$, and moreover, we have that $ \pa_{\tau}\X=\S+ J(\pa_{\sigma} \X) $ from the definition of $\S$. Once again, the fact that $\pa_\sigma \X$ is in $L$ implies that $J(\pa_{\sigma} \X)$ is in $\tilde L$. Notice that $\S$ has to be in $\tilde L$, in general, because, otherwise, the general metric induced on $L$ would not be arbitrary, as follows from $(\nabla_{\S}H_{BC})(\partial_{\sigma}\X)^{B} (\partial_{\sigma}\X)^{C} =0$, which is in turn implied by the null nature of $\S$ and the string EOM. Therefore, in general, $\pa_\tau \X$ has to be in $\tilde L$.
This naturally generalizes the flat case, and it also implies that the usual concept of spacetime metric is associated with the induced metric on $L$,  that is, $H|_L $, a part of a much richer structure of the dynamical phase space description which also includes dynamical momentum space associated with $\tilde L$. 


{\it Conclusion:} We have presented a new viewpoint on string theory, with wide ramifications and applications ranging from the stringy uncertainty principle  \cite{grossm, sup, witten} to ``non-compact'' T-duality  \cite{aldaym}, including the vacuum problem in string theory. Our main point is: {\it The fundamental symmetry of string theory contains diffeomorphisms in phase space. In this formulation both elements $(\eta,J)$ of the chiral structure  are dynamical. The solutions are labelled by bi-Lagrangians and spacetime is a derived dynamical concept.} The fundamental mathematical structure is encoded in the new concept of Born geometry and the choice of bi-Lagrangian structure and the induced metrics on space-time $L$ as well as on momentum space $\tilde{L}$.  
This manifestly implements Born reciprocity and it implies a dynamical, curved phase space, including a dynamical, curved momentum space \cite{dms}, thus providing a generalization of locality.
We note that this formulation can be consistently quantized \cite{flmlong1}.
The implementation of conformal invariance is non-trivial in general, particularly in the interacting case. This is the problem of finding consistent, conformally invariant, string backgrounds. In general, apart from just Weyl invariance we have to enforce worldsheet Lorentz invariance \cite{weyllorentz}. The combination of the two are now required for consistency. We have evidence at one loop (but not yet at all loops) that consistent backgrounds exist, that are not obviously the same as traditional string
backgrounds \cite{thanks}.


\begin{thebibliography}{100}
\bibitem{bornstring}
Note that the usual Hamiltonian and diffeomorphism constraints of string theory in flat target space already implement Born reciprocity!
\bibitem{nb}
In our new formulation of string theory
we do not have to restrict $M$ to be either compact or spacelike; in particular, we can allow compactifications on light-like directions.
\bibitem{footnote1}
Our conventions are such that in the conformal frame the 2d metric is $-d\tau^{2} + \rd \sigma^{2}$, $*\rd\sigma =\rd \sigma$, $*\rd \tau=\rd \tau$ and $\rd\sigma \wedge \rd \tau = \rd^{2}\sigma$.
\bibitem{signature}
Here neutral means that $\eta$ is of signature $(d,d)$, while $H$ is of signature $(2,2(d-2) )$.
\bibitem{chiral}
The Tseytlin action is chiral, and this, roughly speaking, accounts for cutting in half what naively appears as the ``doubling'' of the target space degrees of freedom.
\end{thebibliography}
\end{document}